# Inevitable-Metaverse: A Novel Twitter Dataset for Public Sentiments on Metaverse

Kadhim Hayawi[1*], Sakib Shahriar[1], Mohamed Adel Serhani[2], and Eiman Alothali[2]

[1]College of Technological Innovation, Zayed university, Abu Dhabi, UAE
[2]College of Information Technology, UAE university, Abu Dhabi, UAE

**Abstract.** Metaverse has emerged as a novel technology with the objective to merge the physical world into the virtual world. This technology has seen a lot of interest and investment in recent times from prominent organizations including Facebook which has changed its company name to Meta with the goal of being the leader in developing this technology. Although people in general are excited about the prospects of metaverse due to potential use cases such as virtual meetings and virtual learning environments, there are also concerns due to potential negative consequences. For instance, people are concerned about their data privacy as well as spending a lot of their time on the metaverse leading to negative impacts in real life. Therefore, this research aims to further investigate the public sentiments regarding metaverse on social media. A total of 86565 metaverse-related tweets were used to perform lexicon-based sentiment analysis. Furthermore, various machine and deep learning models with various text features were utilized to predict the sentiment class. The BERT transformer model was demonstrated to be the best at predicting the sentiment categories with 92.6% accuracy and 0.91 F-measure on the test dataset. Finally, the implications and future research directions were also discussed.

---

* This is to indicate the corresponding author. Email address: abdul.hayawi@zu.ac

# 1 Introduction

The era of smartphones and social media has pushed the digital revolution to an unprecedented level. With the availability of the internet and the accessibility of smartphones and other smart devices, millions around the globe spend much of their time online. From obtaining online education to running businesses on social media, the internet has provided a unique opportunity for many. More recently, the metaverse has emerged as a technology and gained significant attention. In particular, with the rebranding of social media giant Facebook into Meta, metaverse has become a trending topic of discussion. By using virtual and augmented realities, metaverse aims to connect people worldwide into an immersive digital experience. While the metaverse has the potential to contribute to beneficial applications, such as medical training and rescue operations, there are notable concerns. For example, the metaverse may be used for cyberbullying as well as identity theft and privacy attacks. Therefore, it is necessary to understand the public sentiment about the acceptance of an up-and-coming technology.

Digital media statistics from 2022 show that 59% of the global population uses social media, with an average daily usage of 2 hours and 29 minutes[1]. Social media platforms can, therefore, be perceived as a 'global town hall' enabling people all around the globe to connect and discuss emerging issues. The increasing adoption of social media has also prompted businesses and governments to perform data-driven analyses of public opinion [1]. Unlike the traditional form of sentiment analysis such as polls and interviews, social media allows researchers with a more accessible method, encompassing a diverse set of audiences. Consequently, researchers have used social media to analyze public sentiment on various topics, including vaccine hesitancy [2], the stock market [3], and election predictions [4]. With the advancement in natural language processing tools and machine learning, processing and analyzing social media posts have become more convenient. It is also possible to extract other insights about a given post, such as the number of views or impressions and popularity or likes. There are also some notable challenges in social media sentiment analysis, including the detection of sarcasm and slang language. Nonetheless, social media remains a suitable form for social media sentiment analysis in the case of metaverse given that a large portion of metaverse adopters will likely be social media users.

Although the rebranding of Facebook to Meta attracted global attention, it is unclear what the general perception regarding the metaverse is. Therefore, this research aims to investigate the public sentiment about Metaverse using social media data. Following are the main contributions of this paper:

- It investigates lexicon-based sentiment analysis on metaverse-related posts on Twitter using natural language processing.
- It introduces a novel metaverse-related dataset containing sentiment scores and sentiment class. The dataset is made available publicly and can be accessed: (https://github.com/SakibShahriar95/Metaverse).
- It utilizes several machine learning and deep learning models to perform metaverse sentiment classification.
- It evaluates and compares the impact of various features and models on metaverse prediction performance.

The rest of the paper is organized as follows. Section 2 provides a background of relevant technology, the evolution of metaverse, and a concise review of related works in the literature. Section 3 describes the overall methods used in the study, including data collection, data processing, sentiment analysis, and machine learning models. Section 4 presents a discussion on the results, comparison, implications, and limitations of the proposed work. Finally, Section 5 concludes this paper.

[1] https://datareportal.com/reports/digital-2022-july-global-statshot

## 2 Literature Review

This section discusses the role of virtual and augmented reality in the realization of the metaverse. Moreover, a historical overview of the development of the metaverse as a technology is presented. Lastly, a concise review of relevant related works in the literature is discussed.

### 2.1 Virtual and Augmented Reality

The interaction between the real or physical world with the digital or virtual world resulted in technologies like augmented reality (AR). The objective of AR technology is not to bring users into a virtual world. Rather, AR can enhance physical world experiences with virtual information. Having real-time information about products projected as a user is browsing through a supermarket is one of the simpler use cases of AR. Virtual reality (VR) on the other hand aims to completely immerse the user into a curated reality, which may mimic real-world experiences but could also be designed to be fictional. There are numerous applications of VR including engaging older adults living in residential aged care facilities who may suffer from isolation [5]. The relationship between AR and VR, which also results in mixed reality (MR) is illustrated in Fig. 1. using the reality-virtuality continuum proposed by Milgram et al [6]. The authors also define MR as an environment consisting of both real world and virtual world objects simultaneously. MR is expected to play an important role in learning environments where a VR headset can be utilized to obtain instructions that can be used to perform tasks on a real environment. In this context, the metaverse provides an environment that can merge physical reality with digital virtuality [7] by providing AR and VR spaces or a combination of the two. The most anticipated use case of the metaverse is a completely immersive VR social media experience where users can put on their VR headsets and transform them into their avatars within the virtual universe. The users can then immerse themselves in a shared experience with their friends as they navigate through the virtual universe. However, the metaverse can also provide various AR experiences including meetings where other participants can be projected onto the real world to enhance the virtual meeting experiences.

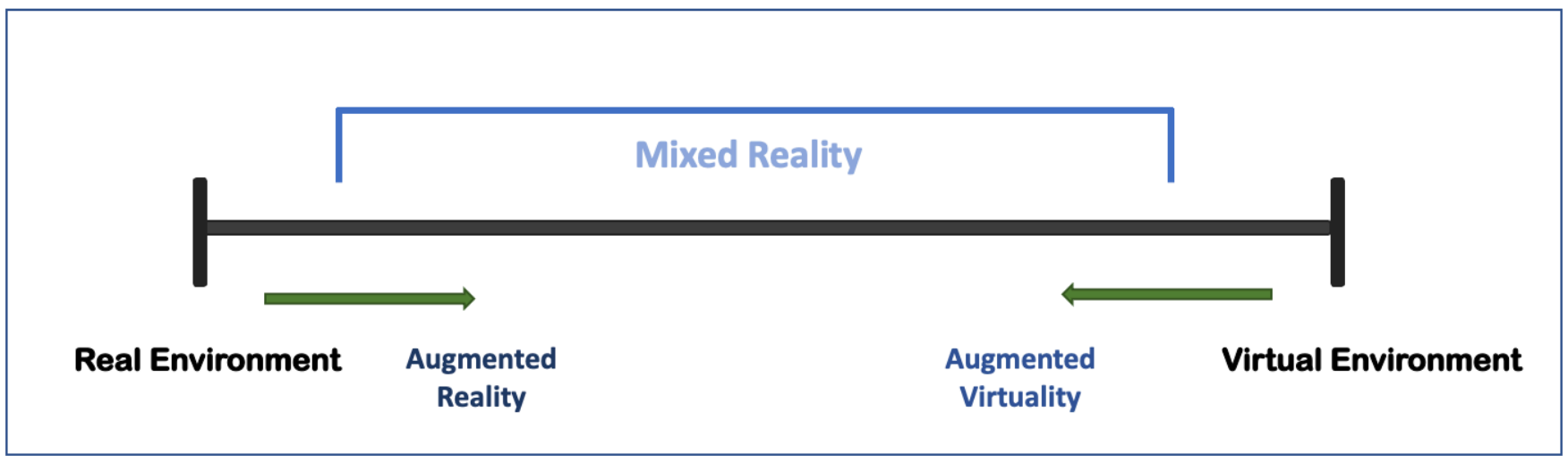

**Fig. 1.** Reality-virtuality continuum [6]

### 2.2 Evolution of the Metaverse

Despite recently emerging as a buzzword, the metaverse finds its origin back in the 1990s. Fig. 2. depicts the timeline and evolution of various key technologies and milestones enabling the metaverse. In August 1991, the first website created by Tim Berners-Lee went live, marking the beginning of internet webpages. Internet as a technology has come leaps and bounds since the early 1990s and is a fundamental technology enabling the metaverse as it allows billions of users to connect online. Metaverse has become a buzzword in the technological sector in recent times. However, the term metaverse was first introduced in 1992 by Neal Stephenson in his science fiction novel Snow Crash. The novel portrayed a 3-dimensional virtual world populated by user-controlled avatars. Blockchain technology is considered to be an important component of the metaverse. The proof of work concept which is used to verify blockchain technology was published in 1993 by Cynthia Dwork and Moni Naor. In 1994, the video game company SEGA released its virtual reality 1 (VR1) amusement park offering an unprecedented experience to users despite being a very simplistic technology compared to its modern counterparts [8]. The beginning of the 21st century was marked by the introduction of Second life, a virtual world platform developed by Linden Lab. This platform provided a virtual experience by allowing social networking and information engagement but ultimately did not provide a pleasant experience due to network bandwidth constraints. In 2006, a gaming platform called Roblox was launched that allows users to create and play games developed by other users. Roblox provides users with an immersive experience containing avatars which enables a 'social hangout' experience. The next crucial technology enabling the metaverse was the introduction of Bitcoin, the world's first cryptocurrency. It is anticipated that cryptocurrencies will be the main source of transactions in the metaverse. Likewise, non-fungible tokens (NFTs) are also integral to the metaverse experience because of their ability to provide digital ownership over items and other valuable assets. Lands and properties owned by users in the metaverse can be verified by NFTs. The focus on developing VR experiences was strengthened with social media giant Facebook acquiring the VR hardware and software provider Oculus in 2014 for 2 billion US dollars [9]. In the following year, the first iteration of decentraland was developed to allow the shared experience of the virtual world. In 2016, Pokémon Go, an AR smartphone game, gained immense popularity. The game used mobile phone GPS coordinates to enable users to locate and capture Pokémon characters as they explored real-world spaces like parks and urban areas [10]. In 2018, another important VR game known as Axie infinity was launched. This game was developed on the Ethereum blockchain and is among the most popular play-to-earn games, where players can collect NFTs and later trade them for cryptocurrencies. Due to the COVID-19 pandemic, a global lockdown was imposed throughout 2020 restricting the movement of people [11]. During this period, metaverse as a concept gained significant interest as it became a pathway for shared experiences and virtual meetings. In April 2020, the first metaverse concert was held by the American rapper Travis Scott on Fortnite, an online videogame platform. The following year Microsoft introduced Mesh, a platform for virtual meetings with mixed reality. In October 2021, Facebook changed its name to Meta signaling its intent on building the metaverse technology.

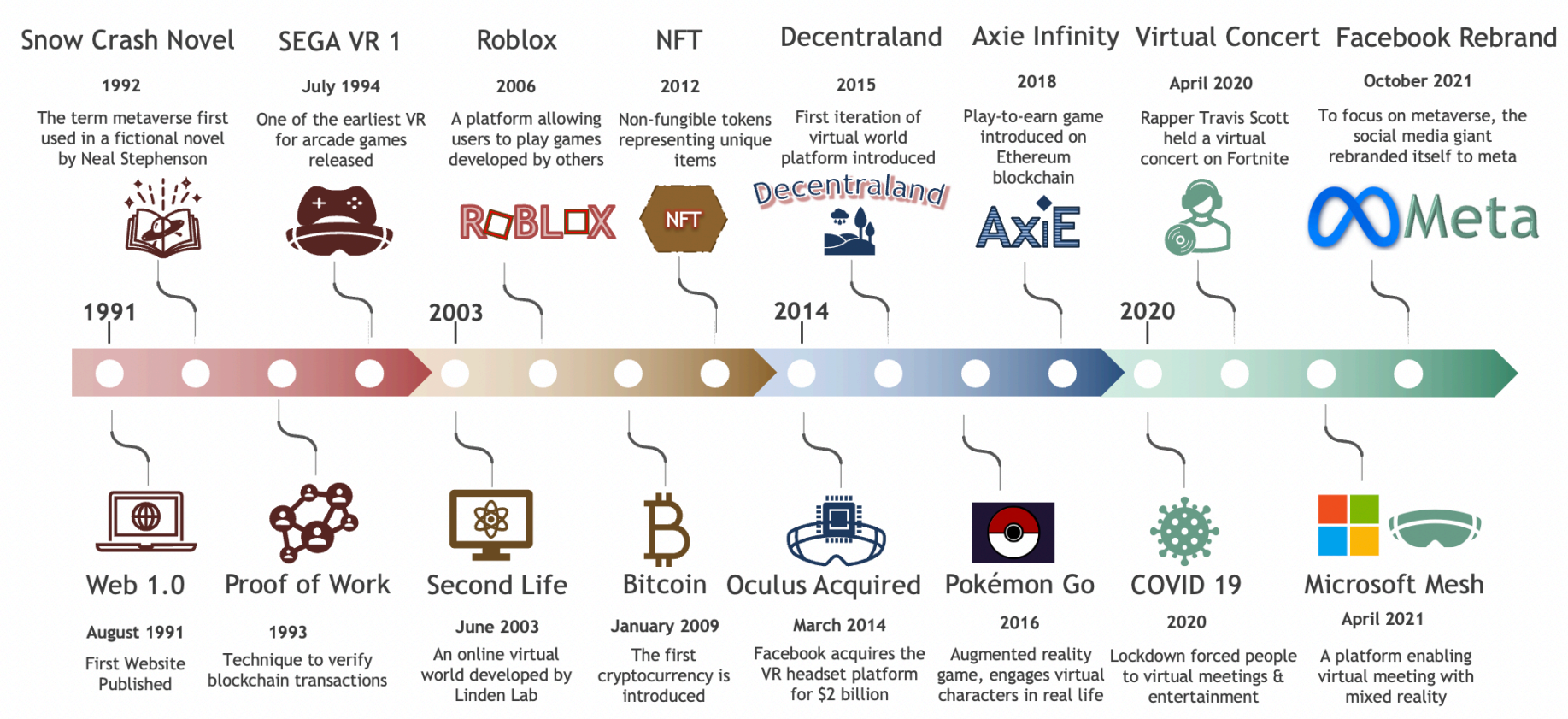

**Fig. 2.** Evolution of the metaverse

## 2.3 Related Works

As the metaverse is currently in the development phase and the research area is still growing, there is limited literature currently available on the metaverse. In this section, some of the relevant metaverse-related works are discussed. Metaverse has the potential to transform the way advertisements are presented to people and how they engage with them. Various research agendas in the context of interactive engagement in the metaverse were proposed by [12]. Some of the research agendas include developing key performance indicators (KPIs) to assess advertising impact as well as addressing advertisement privacy concerns in the metaverse. In [13], the authors discussed the potential impacts on the hospitality and tourism sector with the advent of the metaverse.

The authors argue that unprecedented situations such as the COVID-19 pandemic and the war in Ukraine pose a threat to the hospitality and tourism industries. The metaverse can be a great alternative solution for people to experience activities such as virtual flights, concerts, and skydiving activities. Consequently, businesses will need to adapt to metaverse technologies and software to provide competitive experiences. The authors also highlight the need for investigating the cultural values of users and cross-cultural challenges that may impact their decisions.

As an emerging technology, it is still unclear whether the metaverse will be widely accepted by the people. Aburbeian et al. [14] attempted to gauge the acceptance of metaverse using the technology acceptance model. They conducted a case study with 302 participants who were asked to answer an online Likert scale survey. The authors concluded that users are likely to engage with the metaverse if they found it to be useful and easy to use. The study also found that price negatively impacts user intention suggesting that the cost of metaverse components such as VR headsets must be low enough for users to adopt them. Although the objective in our work is similar in the sense that we want to analyze and predict people's sentiments regarding the metaverse, the methods used are different. Instead of a survey, the focus of our work is to use data visualization and machine learning to analyze and predict metaverse sentiment. Ahmad and Gata [15] analyzed the sentiment of Indonesian people towards the metaverse. They found that majority (66%) of the people showed neutral sentiment towards the technology and the remaining population were split between positive (16%) and negative (17%). By using a linear support vector machine, they obtained an accuracy of 87% in predicting the sentiment. The proposed work in this paper extends the previous work by performing sentiment analysis of English tweets and using both machine and deep learning algorithms. Tunca et al. [16] performed a thematic content analysis of metaverse-related news articles published on The Guardian website. Their sentiment analysis revealed that 61% of the articles were positive, 30% were negative, and the remaining 9% were neutral. Although this study provides a significant contribution to metaverse acceptance, it is limited to news articles. Researchers also analyzed social media sentiment regarding the metaverse a week before and after the announcement of Facebook rebranding itself to Meta [17]. Their dataset was constructed by retrieving all posts containing the hashtag 'metaverse'. Similarly, the authors in [18] conducted sentiment analysis of metaverse by selecting about 5500 popular social media posts, i.e., posts with high impression rates. In contrast, the proposed work introduces a more comprehensive dataset that uses the keyword 'metaverse' covering social media posts over three years.

## 3 Methods

The graphical overview of the proposed application is presented in Fig.3. The details of the implementation are presented next.

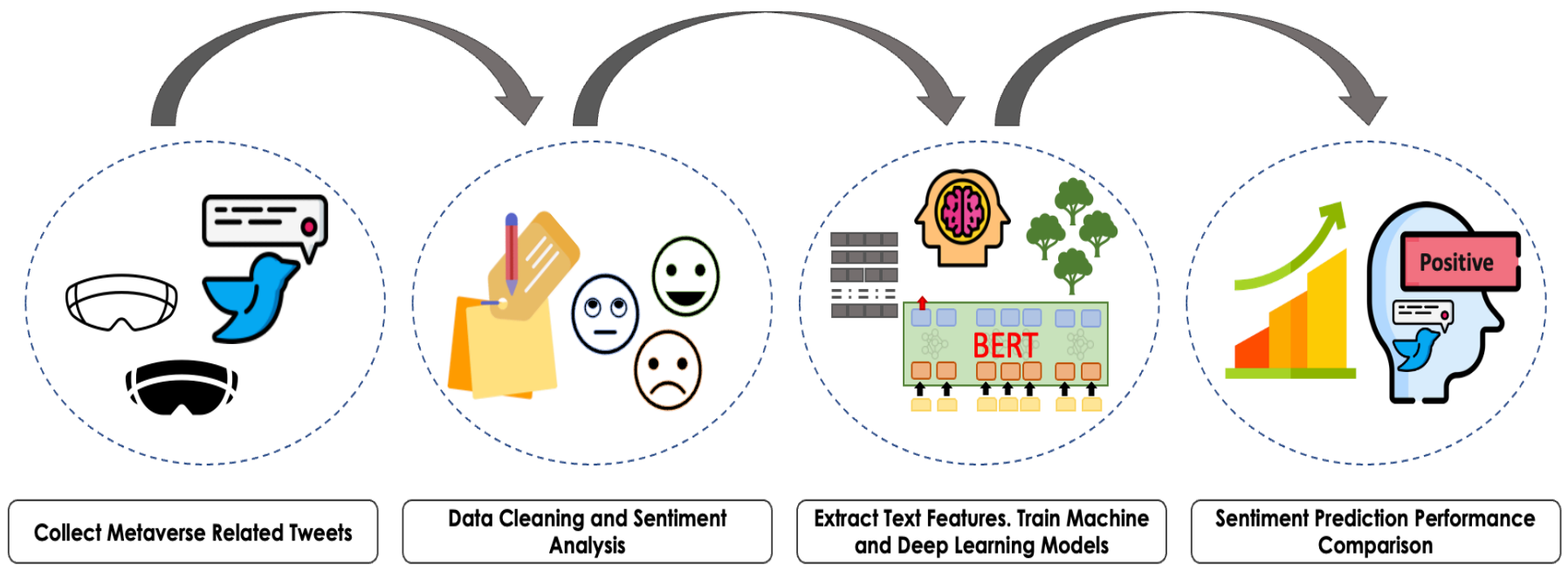

**Fig. 3.** Graphical representation of the proposed application

### 3.1 Dataset Collection

To understand the public sentiment on metaverse, we decided to collect social media posts on this subject. Particularly, we decided to utilize posts on Twitter which consists of 211 million daily active users [19]. To extract posts from Twitter, also called tweets, one can use the Twitter API to obtain the text of the post, user information such as location (if available), and retweets in JSON format. We obtained all relevant tweets by accessing the Twitter API using the *Twarc* library in python and searching with the keyword 'metaverse'. The start date of the search was set to January 1, 2019, and the end date was until and including April 9, 2022. Only tweets in the English language were considered in the search and any retweets, replies, tweets with media, and tweets with links were excluded. The total number of tweets for each year is summarized in Table 1.

**Table 1.** Number of metaverse tweets by year

| Year | Total Tweets |
|---|---|
| 2019 | 3827 |
| 2020 | 4233 |
| 2021 | 344, 887 |
| 2022 (until April 9, 2022) | 396, 895 |
| Total | 749, 842 |

The weekly average metaverse tweet activity, which has been adjusted on a logarithmic scale, is displayed in Fig. 4. During 2019 and 2020, there were very few social media posts regarding metaverse with a maximum of 53 tweets for the first week of October 2019. At the beginning of 2021, we notice that the trend starts to shift as the number of tweets spikes to 209 during the second week of April 2021. The activity keeps on increasing throughout 2021 and early 2022 with the peak recorded during the second week of January 2022 which resulted on average 8247 tweets.

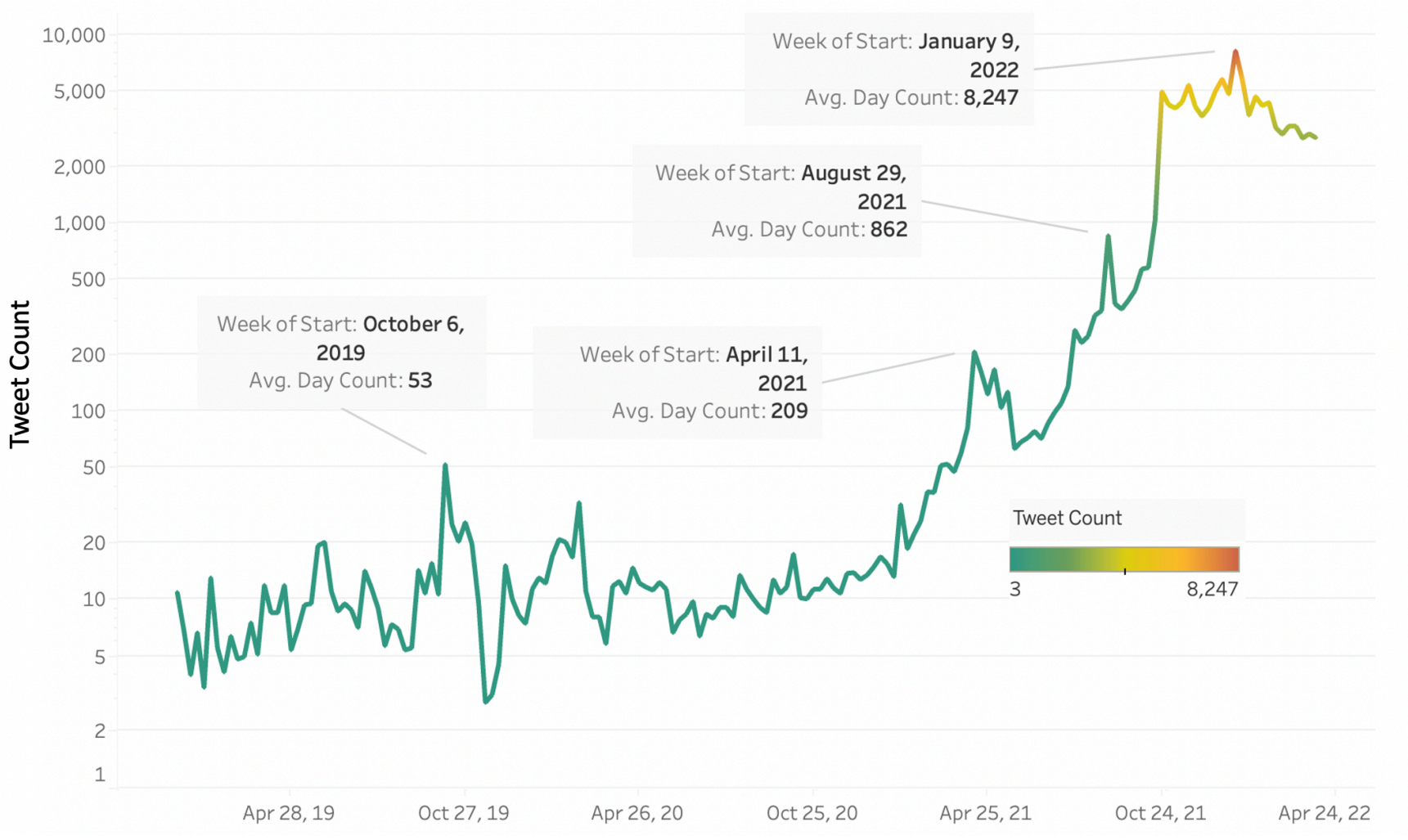

**Fig. 4.** Metaverse tweet frequency with time

## 3.2 Data Preprocessing

Before performing sentiment analysis, it is essential to preprocess the text data adequately. Given that the number of tweets in the dataset was nearly three-quarters of a million (750, 000), it would have required a significantly longer time to compute the sentiment and train the machine learning models with such a large dataset. Consequently, it was decided that from each year, a random sample of 15% of the original dataset was taken for the analysis and predictions. This sampling would ensure all the years in the dataset were represented in the analysis such that there was no bias in sampling. A sample size of 15% also allowed us to obtain a dataset size that would be large enough but not be computationally expensive. The dataset size after sampling resulted in 112, 461 rows of metaverse-related tweets. Furthermore, fields including author location and tweet id were not required and therefore removed. Moreover, punctuation marks, hashtags, external links, and brackets were removed as part of the preprocessing steps. Non-alphabets were also removed, and the texts were all converted to lower case. Common stop words in the English language including 'to', 'is', 'but', and 'has' were removed as these are low information words providing little to no contextual knowledge. Removing these stop words reduces the dataset size and consequently reduces the training complexity. The natural language tool kit library (NLTK) [20] was used to obtain the list of stop words. Finally, any duplicate tweets were removed from the dataset. After completing all the preprocessing steps and removing duplicates, the final dataset contained 86, 565 rows of tweets.

### 3.3 Sentiment Analysis

Sentiment analysis is an approach used to retrieve people's opinions and polarity from written language [21]. To avoid the need to have labeled data, which is time-consuming and expensive, we decided to employ a lexicon-based sentiment analysis (LSA). In LSA, the semantic orientation of words or phrases in a document, which describes the intensity of the words to other words, is analyzed and the sentiment is computed based on the associations of words [22]. A combination of adjectives or adverbs can be used to find the semantic orientation by computing the difference in mutual information between a pre-defined positive and negative word such as 'excellent and 'poor'. Furthermore, [23] demonstrated that the LSA approach can be more advantageous than a classifier approach by simulating the impact of linguistic context. In this work, we utilized the *TextBlob* library [24] in python to perform LSA and obtain the polarity scores. All the words in a sentence were assigned their scores and the final sentiment for a given sentence is calculated by taking the average of all words. The range of the polarity score is between [-1, +1], where -1 represents a completely negative sentiment and +1 represents a completely positive sentiment. Additionally, *TextBlob* also computes the subjectivity of a given sentence and the subjectivity score is between [0, +1]. A higher subjectivity score denotes that the sentence is likely to contain more personal opinions rather than facts. However, in this work, we were mainly interested in obtaining the polarity score and the resultant sentiment based on the polarity score. A tweet is labeled as a negative sentiment if the polarity score is less than 0 and it is labeled as a positive sentiment if the polarity score is more than 0. If the polarity score is equal to 0, then it is labeled as a neutral sentiment. A sample of metaverse-related tweets along with their polarity scores and sentiment class are presented in Table 2.

**Table 2.** Sample of various metaverse tweets and their sentiment

| Tweet | Polarity | Sentiment |
|---|---|---|
| good morning everyonenlet keep building the metaverse | +0.700 | Positive |
| did you know that our animal nfts will be able to breed creating a whole new nft that contains the attributes of its parents that just one of the many things possible in our metaverse nn | +0.267 | Positive |
| im going to buy my ancestral village in the metaverse and turn it into a virtual amazon distribution centre | 0.000 | Neutral |
| i am not smoking but is there a digital cigarette company already created for the metaverse | 0.000 | Neutral |
| fb zuck may be stepping down soon this is bad think they struck out on the metaverse thing or didnt sell it right | -0.190 | Negative |
| correct me if im wrong but isnt the metaverse just going to be like an mmorpg where you have to do your job surrounded by irritating npcs if so ill stick with irl interactions please | -0.467 | Negative |

The distribution for each of the three sentiment classes is displayed in Fig. 5. Above half of all tweets (53%) are represented by a positive sentiment, about one-third (33%) of the tweets were neutral, and a small percentage (14%) of the tweets displayed a negative sentiment.

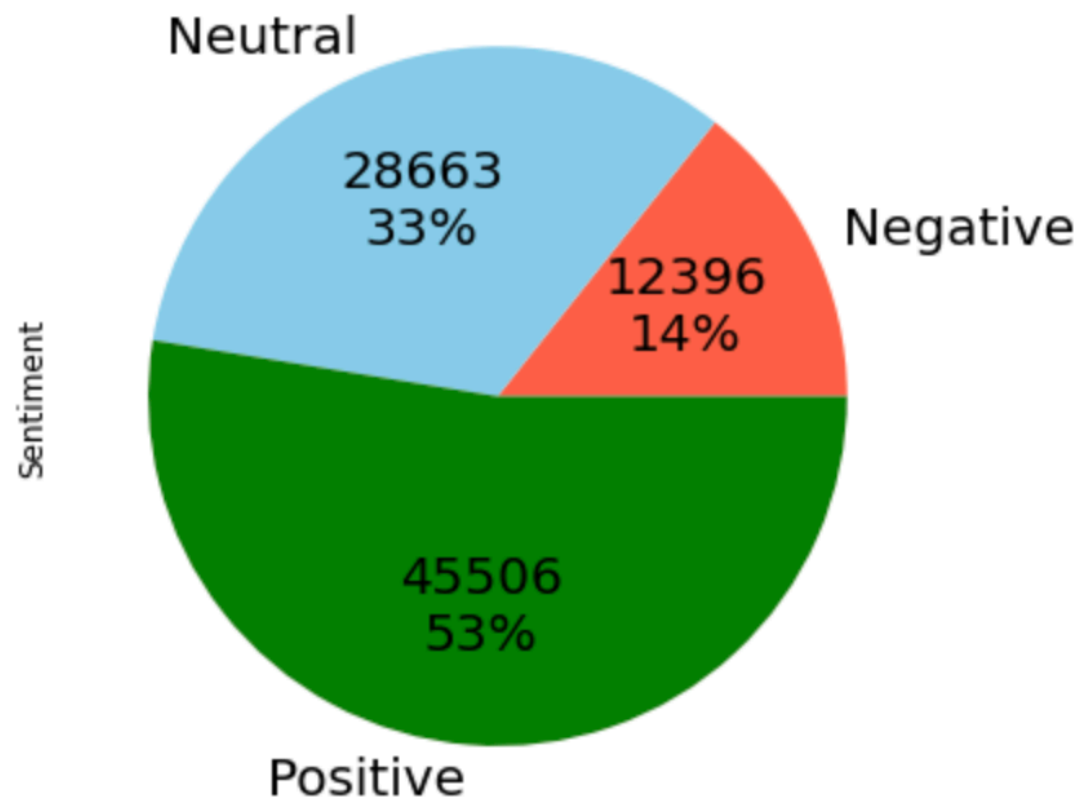

**Fig. 5.** Distribution of the sentiment classes

The world cloud for positive, neutral, and negative metaverse sentiments are illustrated in Fig. 6, Fig. 7, and Fig. 8 respectively. The positive tweets contain a lot of words supporting metaverse adoption including ‘need’, ‘love’, ‘right, ‘future’, and ‘new’. In contrast, the negative tweets besides containing words such as ‘bad’, ‘crazy’, and ‘don’t’, also contained many offensive words. No distinct pattern of words is evident in neutral sentiment tweets. Most tweets belonging to either of the three sentiment classes also contained information about other related technologies such as NFT and cryptocurrency.

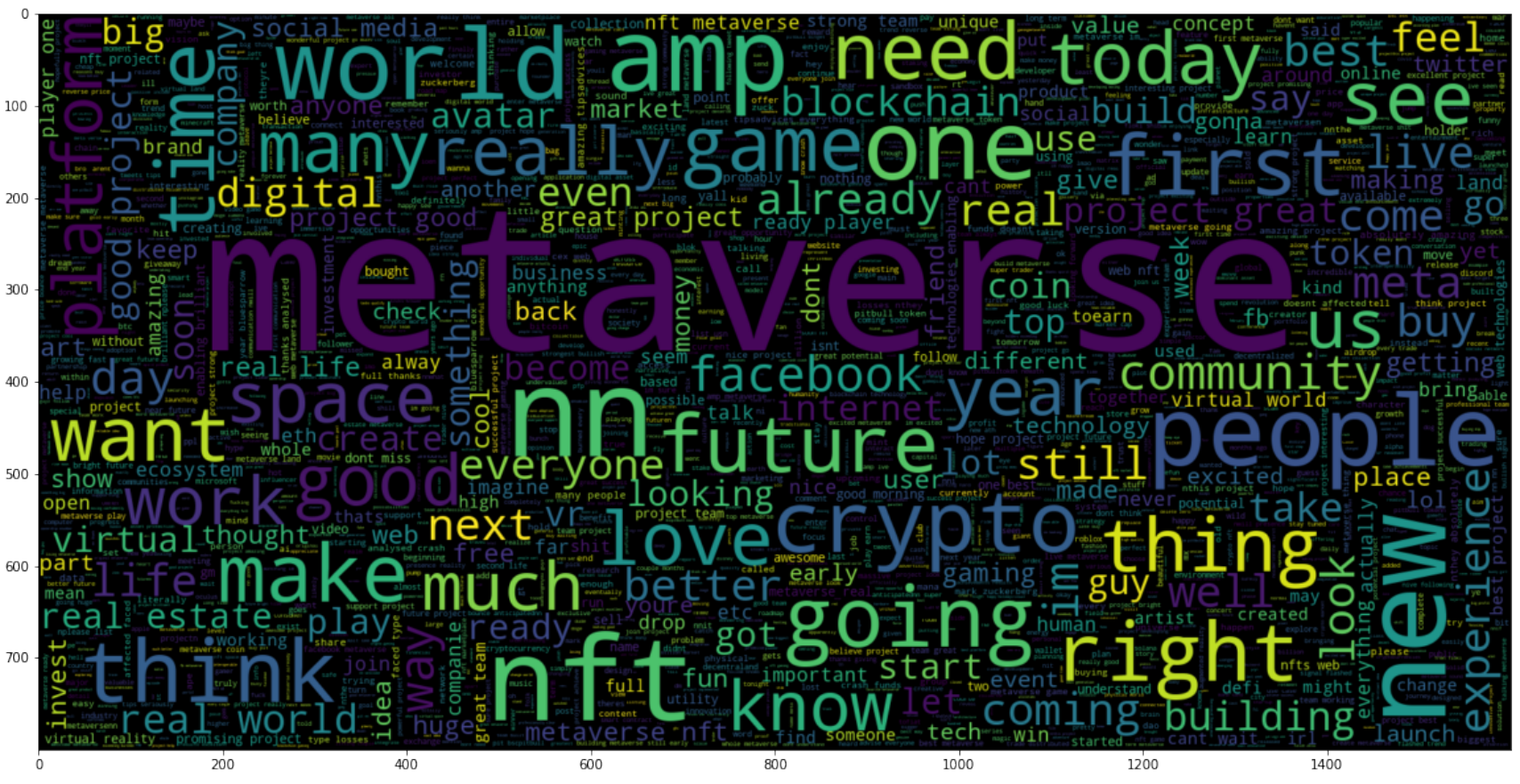

**Fig. 6.** Word cloud visualization for positive sentiment tweets

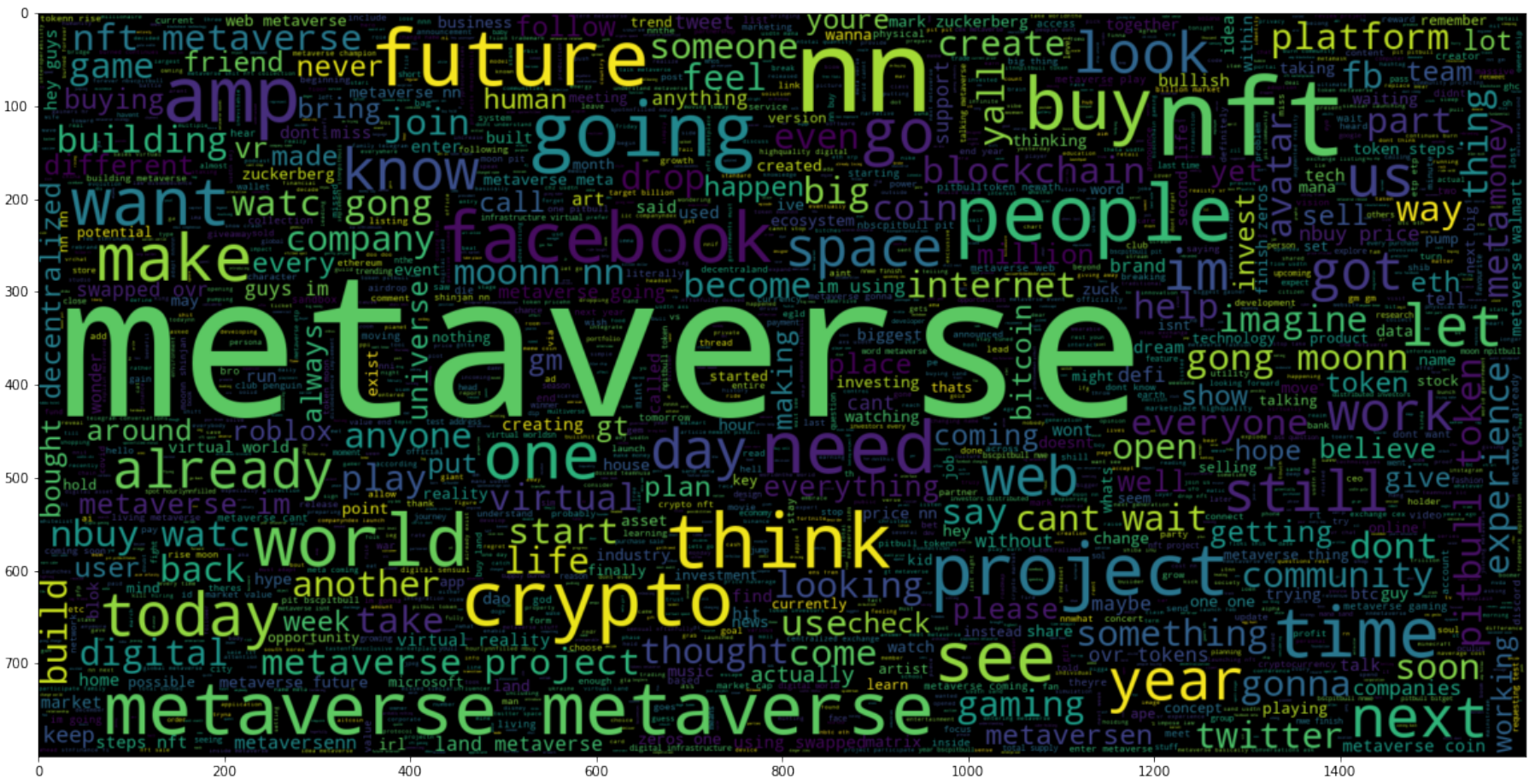

**Fig. 7.** Word cloud visualization for neutral sentiment tweets

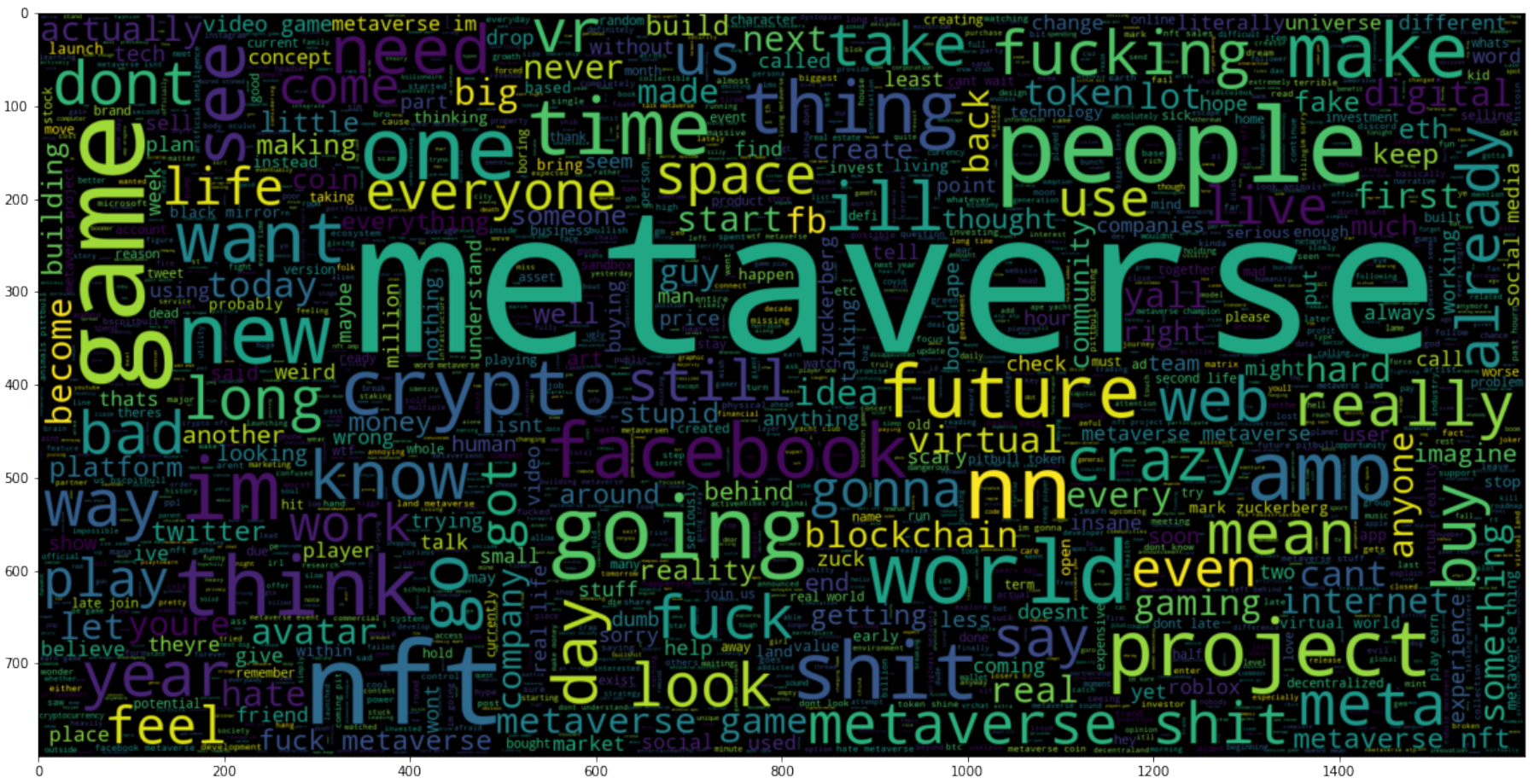

**Fig. 8.** Word cloud visualization for negative sentiment tweets

### 3.4 Machine Learning Models

To train the machine learning models, two features were explored independently, namely bag of words and term frequency-inverse document frequency. Four machine learning algorithms were then trained including a k-nearest neighbor, naïve bayes, random forest, and support vector machine [25].

K-nearest neighbor (K-NN) [26] is a simple learning algorithm that does not require a dedicated training phase. To find out the category of a new data point, a distance measure can be used to compute its k nearest neighbor and the point will be assigned to the category where the majority of the neighbors belong to. Although K-NN is simple, it is prone to outliers in the dataset and does not work well for larger datasets.

Naïve Bayes (NB) [27] is a statistical learning model based on the Bayes theorem. A key assumption of this algorithm is class-conditional independence, i.e., each feature in a class is independent of other features. Although there exist many variations of this algorithm including gaussian NB and Bernoulli NB, in this work we utilized multinomial NB. Multinomial NB is more suitable for text classification problems [28] because the representation of features is generally in terms of word vector counts. NB is computationally inexpensive, but the class-conditional independence assumption does not hold for most real-world applications.

Random forest (RF) is a form of ensemble machine learning where multiple decision trees are aggregated to make predictions. Decision trees are constructed by recursively partitioning the data and each partition is fitted to a simple model [29]. Decision trees are prone to overfitting, and therefore RF overcomes this problem by integrating multiple decision trees and taking their predictions. For classification, majority votes across the decision trees are taken to assign a class for a given data point [30].

Support vector machine (SVM) [31] classifies different data points by constructing a hyperplane (also known as decision boundaries) that can maximize the margin of separation between the different classes in a dataset. To obtain a linear separation, the data points are mapped from a low dimensional to a higher dimensional feature space. Although SVM is generally considered a very accurate algorithm, it requires an extremely lengthy training time, particularly for larger datasets.

A bag of words is considered a very simple approach for creating text features that can be used to train a learning model. First, a list of vocabulary is created from the corpus containing the unique words present in the entire text document. In our case, this is a list of all the unique words present in all the tweets combined. To reduce the complexity, the vocabulary size was limited to include the top 5000 words. The next step is to measure or count the occurrence of each word in individual tweets. The count feature vector is then used for model training. The main disadvantage of this approach is that information regarding the order of the words is not considered and therefore it does not represent a structural context.

The second feature used to train the machine learning models was the term frequency-inverse document frequency (Tf-idf). The values for each word in the document are calculated by the inverse proportion of the frequency of the word in a specific document to the percentage of documents the word appears in [32]. Tf-idf emphasizes the unique words present in each document by giving low importance to words that are common in all the documents. For instance, the word ‘metaverse’ is present in all the tweets, and therefore it does not add much value in discriminating between the different sentiments.

### 3.5 Deep Learning Models

Deep learning algorithms are generally more complex compared to machine learning algorithms and they often outperform the machine learning algorithms when sufficient training data is available. In this work, long short-term memory (LSTM) [33] was used, which is suitable for text and sequential data. The temporal state of the network is preserved by having specialized memory cells and cyclic connections. The LSTM architecture utilized in this work is the one presented in [34]. It consists of a bidirectional LSTM layer with 1-dimensional global max-pooling operation and two dense layers with rectified linear unit (ReLU) activation. We also utilized the Glove word embeddings [35] which convert the words into an n-dimensional space to obtain the semantic similarity between words [34]. To prevent overfitting, we used the dropout layer [36] with a 0.5 rate after each LSTM and dense layers.

Besides LSTM, we also utilized a family of transformer-based deep learning model known as Bidirectional Encoder Representations from Transformers (BERT) [37]. BERT currently provides the state-of-the-art classification on various datasets and is pre-trained on texts from Wikipedia and other English corpora. In contrast to traditional models, BERT looks at text sequences from both directions which enables it to represent the context in language better. The large BERT model was utilized consisting of 24 layers with 1024 hidden dimensions as well as 16 attention heads resulting in 340 million parameters [37]. BERT also has its own tokenizer which was utilized for encoding the input features.

### 3.6 Classification Evaluation

Four popular classification metrics were used to evaluate the performance namely, accuracy, precision, recall, and F-measure. The metrics are defined in Equations (1-4), where TP represents true positive, TN represents true negative, FP represents false positive, and FN represents false negative.

$$Accuracy = \frac{TP + TN}{TP + TN + FP + FN} \quad (1)$$

$$Precision = \frac{TP}{TP + FP} \quad (2)$$

$$Recall = \frac{TP}{FN + TP} \tag{3}$$

$$F - measure = \frac{2 * Precision * Recall}{Precision + Recall} \tag{4}$$

# 4 Results and Discussion

In this section, the results using machine learning and deep learning models for predicting one of the three sentiment categories are compared and discussed.

## 4.1 Machine Learning Results

For each experiment, 75% of the dataset was used to train the model and 25% was used for evaluation. The results of the test set using the four machine learning models for each of the two features are summarized in Table 3.

**Table 3.** Machine learning results om the test set

| | **Bag of Words** | | | | **Tf-idf** | | | |
|---|---|---|---|---|---|---|---|---|
| | Accuracy | Precision | Recall | F-measure | Accuracy | Precision | Recall | F-measure |
| NB | 79.0 | 0.76 | 0.75 | 0.75 | 73.9 | 0.62 | 0.81 | 0.65 |
| KNN | 55.9 | 0.51 | 0.70 | 0.48 | 46.0 | 0.44 | 0.70 | 0.38 |
| RF | 88.1 | 0.82 | 0.87 | 0.84 | **86.8** | **0.80** | **0.87** | **0.82** |
| SVM | **90.9** | **0.87** | **0.89** | **0.88** | 86.5 | 0.80 | 0.86 | 0.81 |

Using Bag of words as input feature, SVM performed the best with 90.9% accuracy and 0.88 F-measure. On the other hand, using Tf-idf as input feature, RF performed the best with 86.8% accuracy and 0.82 F-measure. Fig. 9. and Fig. 10. present the confusion matrix on the test set for SVM with bag of words and random forest with Tf-idf respectively.

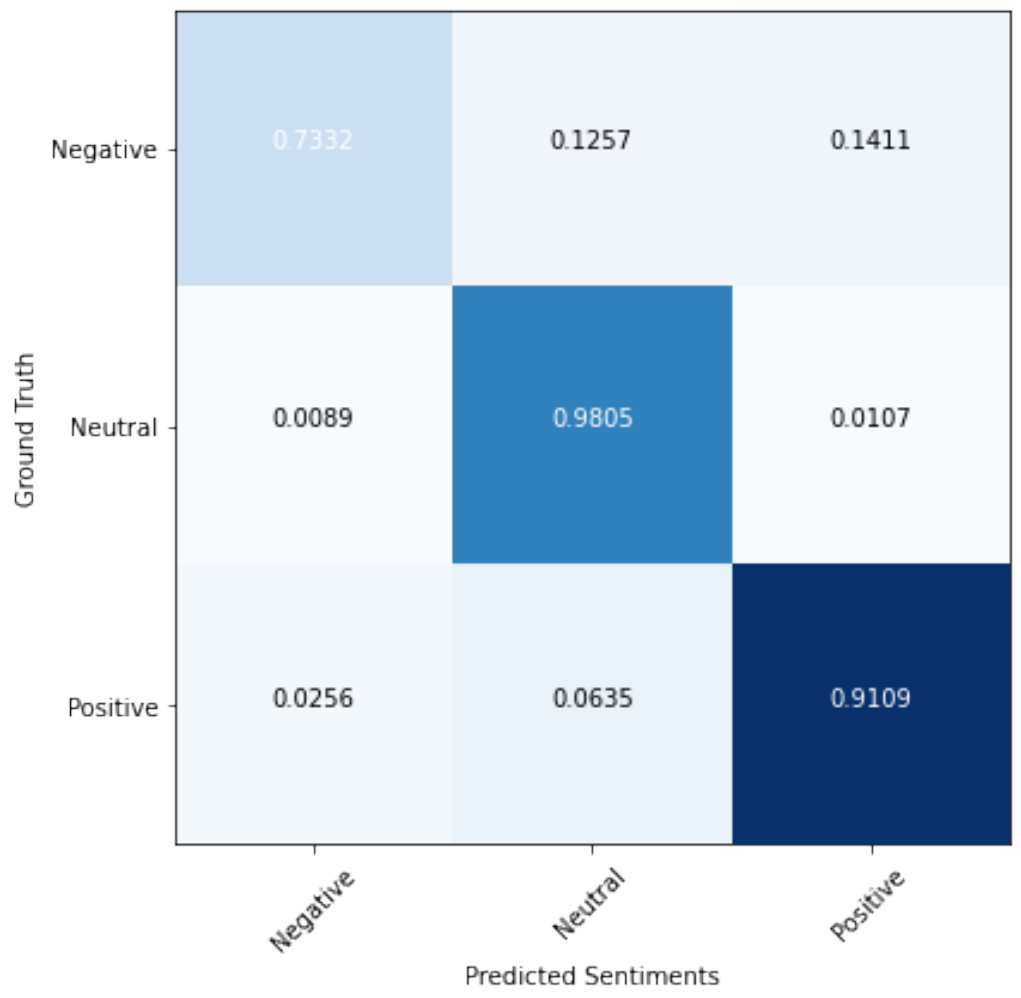

**Fig. 9.** Confusion Matrix on the test set using SVM and bag of words

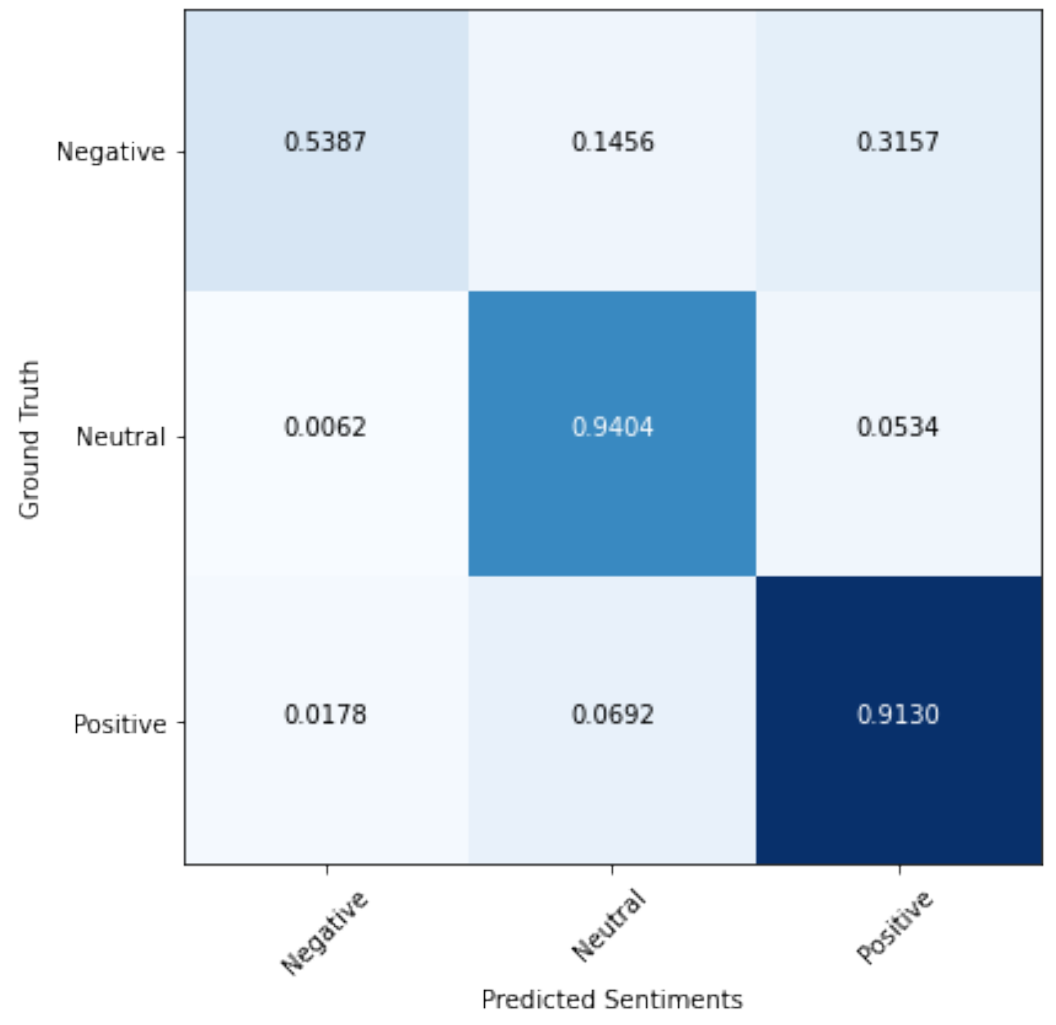

**Fig. 10.** Confusion matrix on the test set using RF and Tf-idf

For both features, it is evident that the models perform the worst in predicting negative sentiment tweets. In the case of bag of words and SVM model, only 0.73 precision was obtained in correctly predicting the negative sentiment with a corresponding F1-measure of 0.79. The performance is even worse using Tf-idf and RF model with 0.54 precision and 0.67 F1-measure. For both models, the best performance was obtained for the neutral sentiment class with the precision of 0.98 and 0.94 respectively.

## 4.2 Deep Learning Results

As with the machine learning models, 75% of the dataset was used for training the deep learning models and 25% was used for evaluation. Fig. 11. and Fig. 12. display the training and validation accuracy curves for LSTM and BERT respectively. These curves were monitored to ensure no overfitting took place during the training phase. Since both the curves are very close to each other, there is no indication of overfitting.

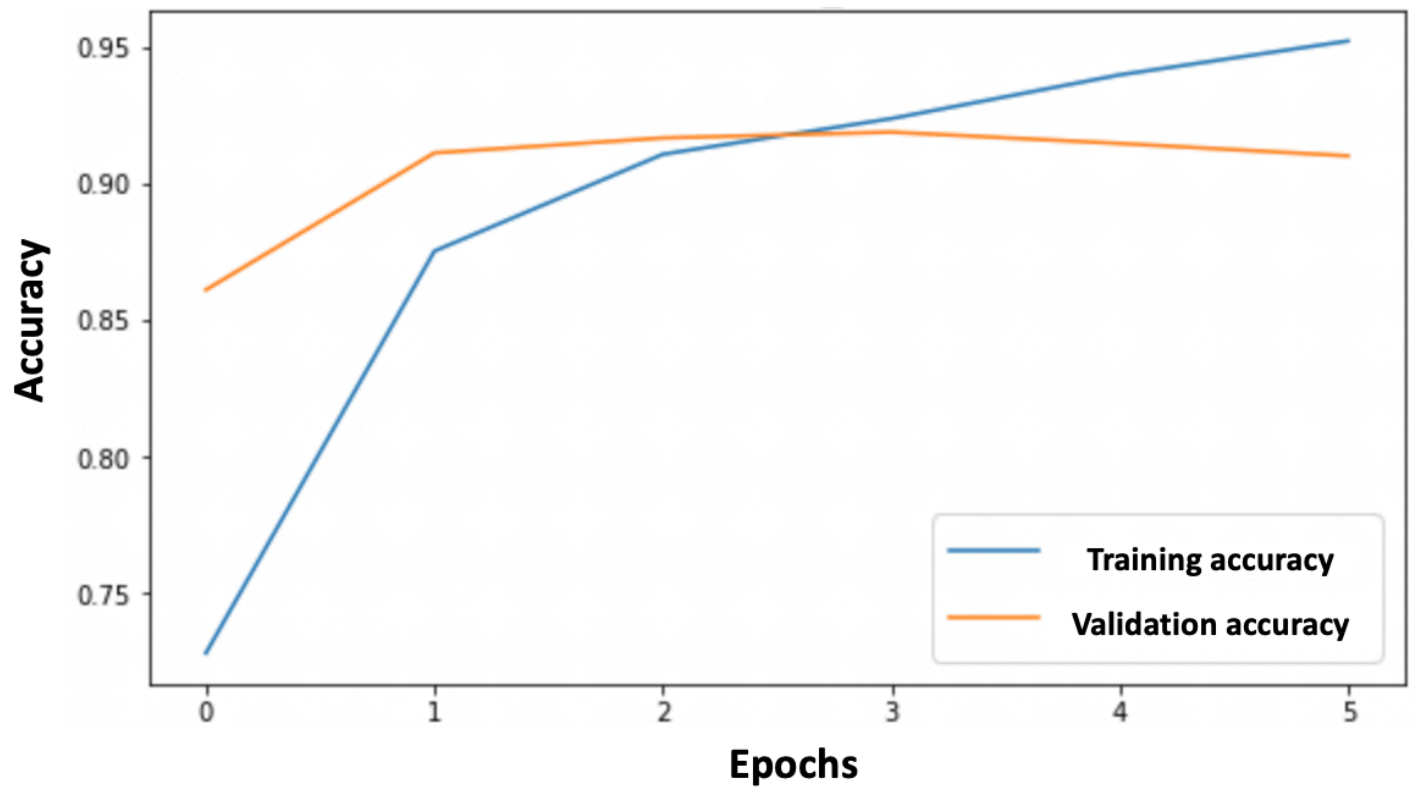

**Fig. 11.** Training and validation accuracies using LSTM

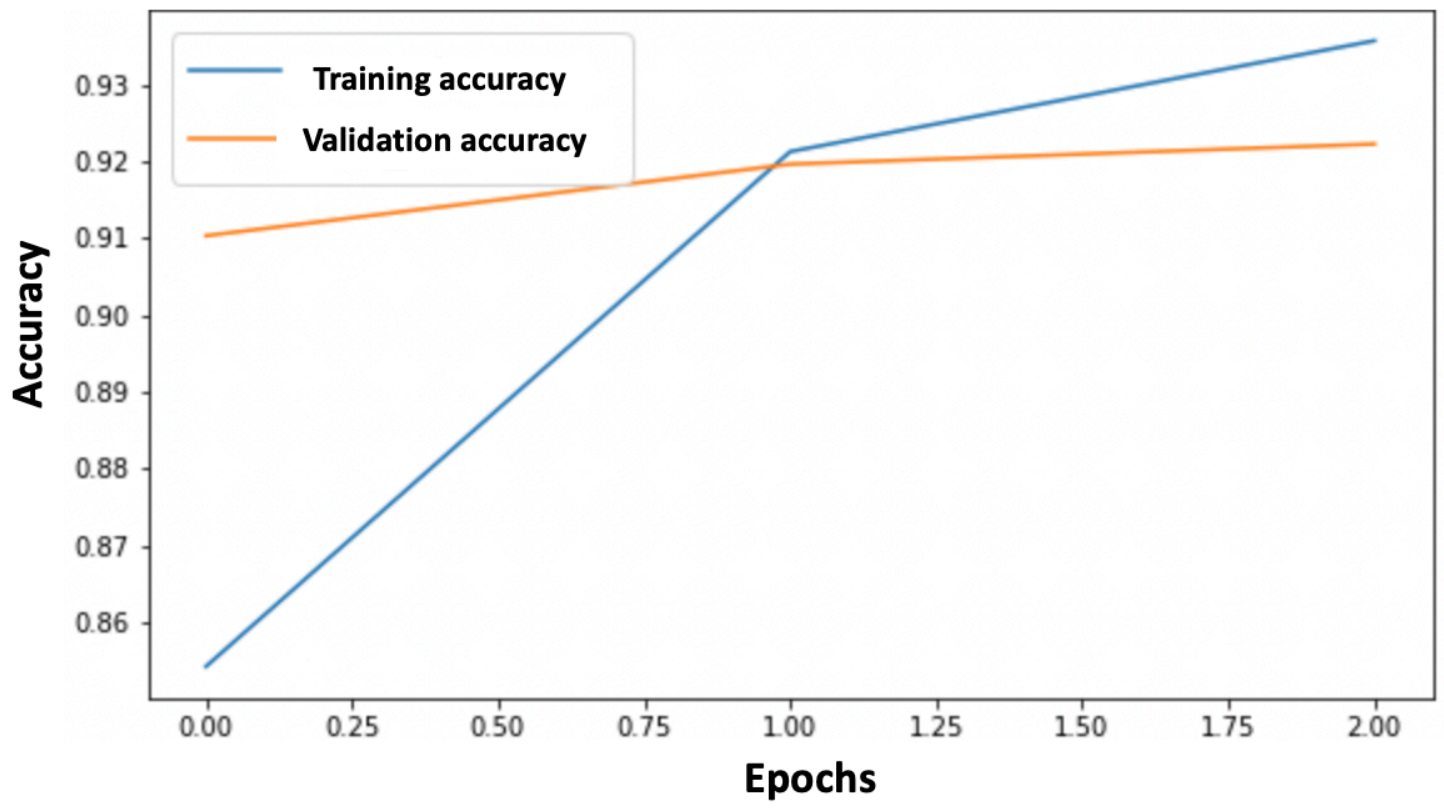

**Fig. 12.** Training and validation accuracies using BERT

LSTM was trained for 5 epochs of the dataset and BERT was trained for 2 epochs of the dataset. The results for both LSTM and the BERT transformer model on the test set are summarized in Table 4.

**Table 4.** Deep learning results on the test set

| | Accuracy | Precision | Recall | F-measure |
|---|---|---|---|---|
| LSTM | 90.7 | 0.89 | 0.89 | 0.89 |
| BERT | **92.6** | **0.90** | **0.91** | **0.91** |

The results indicate that both deep learning models performed better than the traditional machine learning models. The best performance was obtained using the BERT transformer model with 92.6% accuracy and 0.91 F1-measure. Fig. 13. and Fig. 14. present the confusion matrix on the test set for LSTM and BERT respectively.

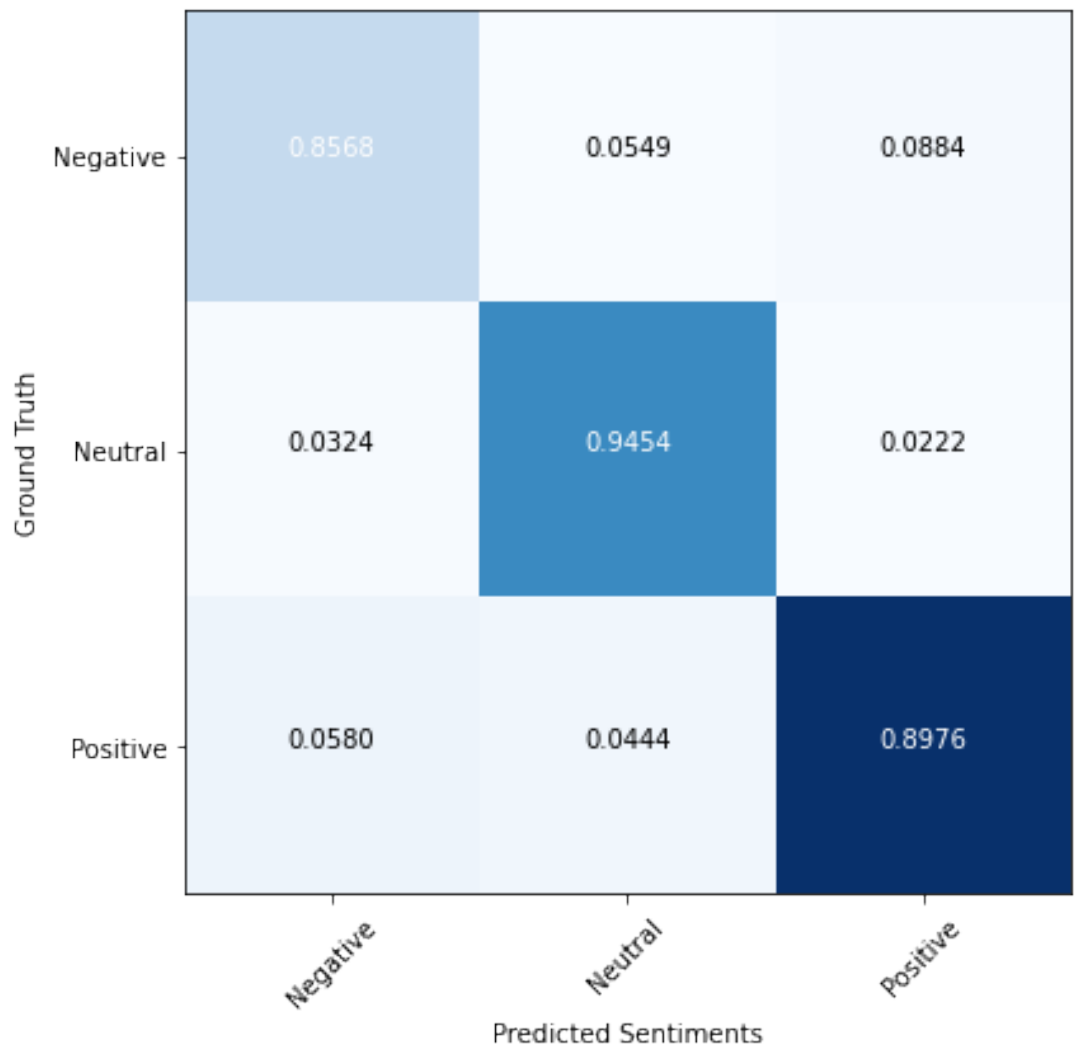

**Fig. 13.** Confusion matrix on the test set using LSTM

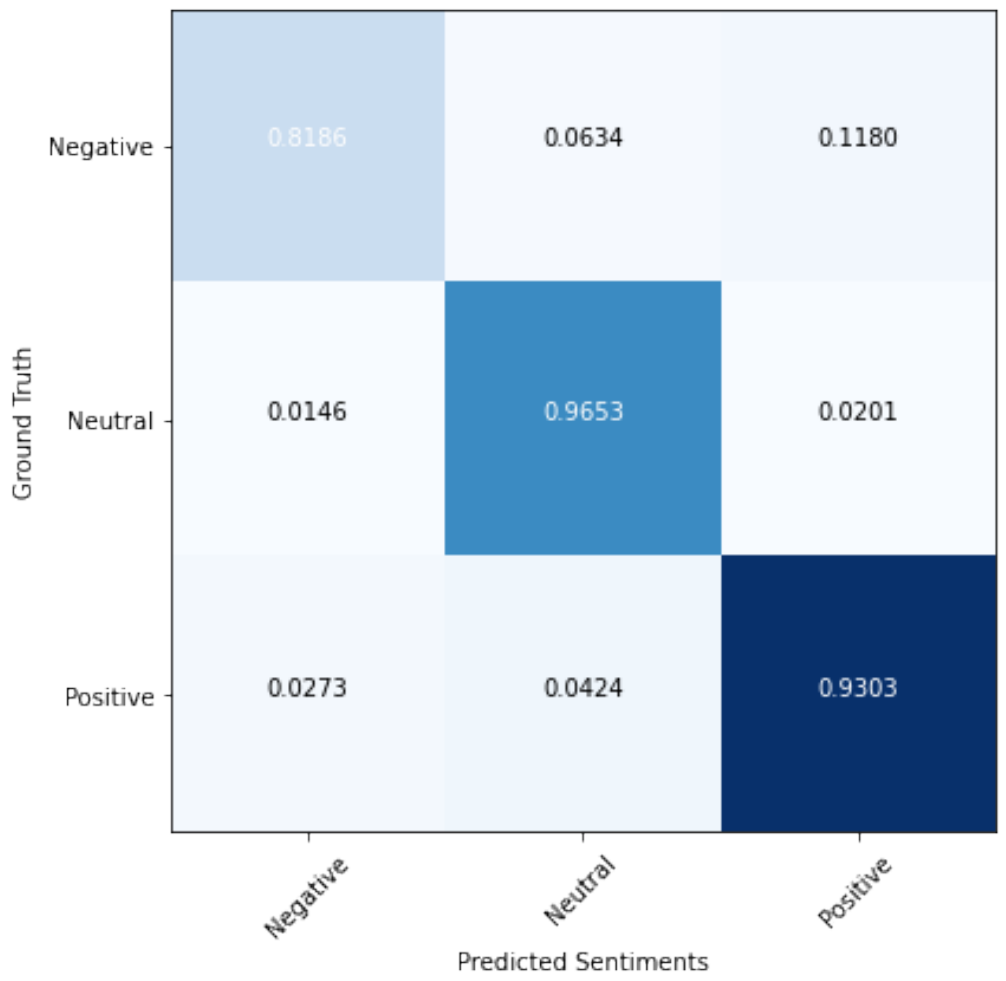

**Fig. 14.** Confusion matrix on the test set using BERT

Both deep learning models can improve upon the machine learning models in predicting negative sentiment. Although BERT performs better overall, LSTM is able to provide the best performance in detecting negative sentiment with 0.86 precision as compared to BERT's 0.82 precision. Moreover, neutral sentiment is the most accurately predicted class for both models.

### 4.3 Comparison

For the machine learning models, better performance was obtained when using bag of words as input feature compared to Tf-idf. Moreover, the two simpler models K-NN and NB failed to provide meaningful predictions using both features. The deep learning models improved upon the machine learning models as expected due to the large dataset size. Across all models, the negative sentiment was least accurately predicted. This is likely due to the negative sentiment being the minority class in the dataset containing only 14% of the samples. However, the neutral sentiment was most accurately predicted despite not being the majority class.

### 4.4 Implications

With growing interest and skepticism surrounding metaverse technology, this research sheds light on the metaverse sentiment on Twitter. Firstly, we observed that the social media activity with regard to the metaverse topic dramatically increased over the last 2 years. After performing lexicon-based sentiment analysis, we also found that the majority of the metaverse-related tweets displayed positive sentiments and a low number of negative sentiments was present. This offers early promising signs for the adoption of metaverse technology. This research can support organizations and governments to assess the general sentiment surrounding metaverse technology and consequently make their decisions about metaverse adoption.

### 4.5 Limitations

This paper introduced a novel dataset that contains metaverse-related posts on Twitter. A limitation of the dataset is that it was collected using the keyword 'metaverse'. Therefore, relevant tweets containing metaverse-related discussions but not the keyword could have been ignored. For instance, it is likely that many users might have used 'meta' or 'virtual world' in their posts to share their thoughts on the metaverse. In addition, we only considered lexicon-based approach in extracting sentiment of a given post. Despite its advantages, it remains to be seen if other sentiment extraction methods such as VADER [38] would provide a better estimation of public sentiment. In addition, this work is limited by the use of Twitter as the sole social media platform used for sentiment analysis on metaverse. Finally, in this work, our scope was limited to the extraction and classification of three sentiment classes. For a more comprehensive analysis, other sentiment categories, including fear, joy, and excitement, can be extracted using more sophisticated approaches.

### 4.6 Future Work

While this work focused on analyzing public sentiment regarding the metaverse on Twitter, other social media platforms were unexplored. The public opinion on other platforms such as Meta (Facebook) or Reddit may be completely different. Consequently, future research should consider a range of social media platforms as well as offline sources including surveys to provide a more comprehensive analysis of the public sentiment about the metaverse. Moreover, this work is limited by a specific type of sentiment analysis which is lexicon-based. However, deep learning approaches may be superior ([39], [40]) and should be experimented with for future research. Another challenge in social media sentiment analysis is the presence of social media bots that hampers the trustworthiness of a given topic by manipulating public opinion [41]. Consequently, sophisticated methods including the ones proposed by [42] and [43] should be utilized to mitigate the presence of posts generated by social bots. Finally, to better understand the public sentiment regarding the metaverse, attributes such as hashtags can provide better context about a general or a specific trend [44].

## 5 Conclusion

The anticipated arrival of the metaverse technology has generated great hype as well as confusion among internet users. In this work, we briefly discussed the metaverse technology and presented its development over the years. We also collected metaverse-related posts on Twitter to analyze the sentiment related to this technology. Using LSA, we discovered that most of the tweets related to the metaverse are positive and that the negative sentiments are very few. Finally, we also trained four machine learning models and two deep learning models to classify these tweets into the three sentiment categories. The transformer-based BERT model obtained the best performance with an F1-measure of 0.91 on the test set. We also highlighted the potential implications and outlined future research directions.

## Statements and Declarations

**Ethical Approval.** Ethical approval was not required because no personal data was used. Any analysis presented were aggregated.

**Funding.** This work was supported by Zayed University under the research grant RIF R20132.

**Competing Interests.** No potential competing interest was reported by the authors.

**Data Availability.** The dataset is made available publicly and has been anonymized in compliance with Twitter's Terms of service. The dataset contains only the tweet IDs and their respective sentiment scores. The dataset can be accessed using the following link: https://github.com/SakibShahriar95/Metaverse.

## References

[1] O. Belkahla Driss, S. Mellouli, and Z. Trabelsi, “From citizens to government policy-makers: Social media data

analysis," *Gov. Inf. Q.*, vol. 36, no. 3, pp. 560–570, Jul. 2019, doi: 10.1016/j.giq.2019.05.002.
[2] H. Piedrahita-Valdés *et al.*, "Vaccine Hesitancy on Social Media: Sentiment Analysis from June 2011 to April 2019," *Vaccines*, vol. 9, no. 1, Art. no. 1, Jan. 2021, doi: 10.3390/vaccines9010028.
[3] J. Bollen, H. Mao, and X. Zeng, "Twitter mood predicts the stock market," *J. Comput. Sci.*, vol. 2, no. 1, pp. 1–8, Mar. 2011, doi: 10.1016/j.jocs.2010.12.007.
[4] P. Chauhan, N. Sharma, and G. Sikka, "The emergence of social media data and sentiment analysis in election prediction," *J. Ambient Intell. Humaniz. Comput.*, vol. 12, no. 2, pp. 2601–2627, Feb. 2021, doi: 10.1007/s12652-020-02423-y.
[5] S. Baker *et al.*, "Evaluating the use of interactive virtual reality technology with older adults living in residential aged care," *Inf. Process. Manag.*, vol. 57, no. 3, p. 102105, May 2020, doi: 10.1016/j.ipm.2019.102105.
[6] P. Milgram, H. Takemura, A. Utsumi, and F. Kishino, "Augmented reality: a class of displays on the reality-virtuality continuum," in *Telemanipulator and Telepresence Technologies*, Dec. 1995, vol. 2351, pp. 282–292. doi: 10.1117/12.197321.
[7] S. Mystakidis, "Metaverse," *Encyclopedia*, vol. 2, no. 1, Art. no. 1, Mar. 2022, doi: 10.3390/encyclopedia2010031.
[8] "The Virtual Arena – Blast from the Past: The VR-1," *GMW3*. https://www.gmw3.com/2020/07/the-virtual-arena-blast-from-the-past-the-vr-1/ (accessed Jun. 08, 2022).
[9] B. Solomon, "Facebook Buys Oculus, Virtual Reality Gaming Startup, For $2 Billion," *Forbes*. https://www.forbes.com/sites/briansolomon/2014/03/25/facebook-buys-oculus-virtual-reality-gaming-startup-for-2-billion/ (accessed Jun. 09, 2022).
[10] L. J. Dorward, J. C. Mittermeier, C. Sandbrook, and F. Spooner, "Pokémon Go: Benefits, Costs, and Lessons for the Conservation Movement," *Conserv. Lett.*, vol. 10, no. 1, pp. 160–165, 2017, doi: 10.1111/conl.12326.
[11] S. Shahriar and A. R. Al-Ali, "Impacts of COVID-19 on Electric Vehicle Charging Behavior: Data Analytics, Visualization, and Clustering," *Appl. Syst. Innov.*, vol. 5, no. 1, Art. no. 1, Feb. 2022, doi: 10.3390/asi5010012.
[12] J. Kim, "Advertising in the Metaverse: Research Agenda," *J. Interact. Advert.*, vol. 21, no. 3, pp. 141–144, Sep. 2021, doi: 10.1080/15252019.2021.2001273.
[13] D. Gursoy, S. Malodia, and A. Dhir, "The metaverse in the hospitality and tourism industry: An overview of current trends and future research directions," *J. Hosp. Mark. Manag.*, vol. 0, no. 0, pp. 1–8, May 2022, doi: 10.1080/19368623.2022.2072504.
[14] A. M. Aburbeian, A. Y. Owda, and M. Owda, "A Technology Acceptance Model Survey of the Metaverse Prospects," *AI*, vol. 3, no. 2, Art. no. 2, Jun. 2022, doi: 10.3390/ai3020018.
[15] A. Ahmad and W. Gata, "Sentimen Analisis Masyarakat Indonesia di Twitter Terkait Metaverse dengan Algoritma Support Vector Machine," *J. JTIK J. Teknol. Inf. Dan Komun.*, vol. 6, no. 4, Art. no. 4, Mar. 2022, doi: 10.35870/jtik.v6i4.569.
[16] M. S. Tunca, B. Sezen, and V. Wilk, "An Exploratory Content and Sentiment Analysis of The Guardian Metaverse Articles Using Leximancer and Natural Language Processing," 2022.
[17] Ö. Ağrali and Ö. Aydin, "Tweet Classification and Sentiment Analysis on Metaverse Related Messages," *J. Metaverse*, vol. 1, no. 1, Art. no. 1, Dec. 2021.
[18] S. Tunca, B. SEZEN, and Y. S. BALCIOĞLU, "TWITTER ANALYSIS FOR METAVERSE LITERACY', 4," in *INTERNATIONAL NEW YORK ACADEMIC RESEARCH CONGRESS*, 2022.
[19] "100 Social Media Statistics For 2022 [+Infographic] | Statusbrew," *Statusbrew Blog*, Dec. 09, 2021. https://statusbrew.com/insights/social-media-statistics/ (accessed May 21, 2022).
[20] S. Bird and E. Loper, "NLTK: the natural language toolkit," 2004.
[21] B. Liu, "Sentiment analysis and opinion mining," *Synth. Lect. Hum. Lang. Technol.*, vol. 5, no. 1, pp. 1–167, 2012.
[22] P. D. Turney, "Thumbs up or thumbs down? semantic orientation applied to unsupervised classification of reviews," in *Proceedings of the 40th Annual Meeting on Association for Computational Linguistics*, USA, Jul. 2002, pp. 417–424. doi: 10.3115/1073083.1073153.
[23] M. Taboada, J. Brooke, M. Tofiloski, K. Voll, and M. Stede, "Lexicon-based methods for sentiment analysis," *Comput. Linguist.*, vol. 37, no. 2, pp. 267–307, 2011.
[24] S. Loria and others, "textblob Documentation," *Release 015*, vol. 2, p. 269, 2018.
[25] G. Bonaccorso, *Machine learning algorithms*. Packt Publishing Ltd, 2017.
[26] T. Cover and P. Hart, "Nearest neighbor pattern classification," *IEEE Trans. Inf. Theory*, vol. 13, no. 1, pp. 21–27, Jan. 1967, doi: 10.1109/TIT.1967.1053964.
[27] I. Rish and others, "An empirical study of the naive Bayes classifier," in *IJCAI 2001 workshop on empirical methods in artificial intelligence*, 2001, vol. 3, no. 22, pp. 41–46.
[28] A. M. Kibriya, E. Frank, B. Pfahringer, and G. Holmes, "Multinomial Naive Bayes for Text Categorization Revisited," in *AI 2004: Advances in Artificial Intelligence*, Berlin, Heidelberg, 2005, pp. 488–499. doi: 10.1007/978-3-540-30549-1_43.
[29] W.-Y. Loh, "Classification and regression trees," *Wiley Interdiscip. Rev. Data Min. Knowl. Discov.*, vol. 1, no. 1, pp. 14–23, 2011.
[30] A. Liaw, M. Wiener, and others, "Classification and regression by randomForest," *R News*, vol. 2, no. 3, pp. 18–22, 2002.
[31] C. J. C. Burges, "A Tutorial on Support Vector Machines for Pattern Recognition," *Data Min. Knowl. Discov.*, vol. 2, no. 2, pp. 121–167, Jun. 1998, doi: 10.1023/A:1009715923555.
[32] J. Ramos and others, "Using tf-idf to determine word relevance in document queries," in *Proceedings of the first instructional conference on machine learning*, 2003, vol. 242, no. 1, pp. 29–48.

[33] S. Hochreiter and J. Schmidhuber, "Long Short-Term Memory," *Neural Comput.*, vol. 9, no. 8, pp. 1735–1780, Nov. 1997, doi: 10.1162/neco.1997.9.8.1735.

[34] K. Hayawi, S. Shahriar, M. A. Serhani, I. Taleb, and S. S. Mathew, "ANTi-Vax: a novel Twitter dataset for COVID-19 vaccine misinformation detection," *Public Health*, vol. 203, pp. 23–30, Feb. 2022, doi: 10.1016/j.puhe.2021.11.022.

[35] J. Pennington, R. Socher, and C. D. Manning, "Glove: Global vectors for word representation," in *Proceedings of the 2014 conference on empirical methods in natural language processing (EMNLP)*, 2014, pp. 1532–1543.

[36] N. Srivastava, G. Hinton, A. Krizhevsky, I. Sutskever, and R. Salakhutdinov, "Dropout: a simple way to prevent neural networks from overfitting," *J. Mach. Learn. Res.*, vol. 15, no. 1, pp. 1929–1958, 2014.

[37] J. Devlin, M.-W. Chang, K. Lee, and K. Toutanova, "BERT: Pre-training of Deep Bidirectional Transformers for Language Understanding," arXiv, arXiv:1810.04805, May 2019. doi: 10.48550/arXiv.1810.04805.

[38] C. Hutto and E. Gilbert, "VADER: A Parsimonious Rule-Based Model for Sentiment Analysis of Social Media Text," *Proc. Int. AAAI Conf. Web Soc. Media*, vol. 8, no. 1, Art. no. 1, May 2014, doi: 10.1609/icwsm.v8i1.14550.

[39] R. Catelli, S. Pelosi, and M. Esposito, "Lexicon-based vs. Bert-based sentiment analysis: A comparative study in Italian," *Electronics*, vol. 11, no. 3, p. 374, 2022.

[40] H. Zhao, Z. Liu, X. Yao, and Q. Yang, "A machine learning-based sentiment analysis of online product reviews with a novel term weighting and feature selection approach," *Inf. Process. Manag.*, vol. 58, no. 5, p. 102656, Sep. 2021, doi: 10.1016/j.ipm.2021.102656.

[41] M. Ebrahimi, A. H. Yazdavar, and A. Sheth, "Challenges of Sentiment Analysis for Dynamic Events," *IEEE Intell. Syst.*, vol. 32, no. 5, pp. 70–75, Sep. 2017, doi: 10.1109/MIS.2017.3711649.

[42] K. Hayawi, S. Mathew, N. Venugopal, M. M. Masud, and P.-H. Ho, "DeeProBot: a hybrid deep neural network model for social bot detection based on user profile data," *Soc. Netw. Anal. Min.*, vol. 12, no. 1, p. 43, Mar. 2022, doi: 10.1007/s13278-022-00869-w.

[43] E. Alothali, K. Hayawi, and H. Alashwal, "Hybrid feature selection approach to identify optimal features of profile metadata to detect social bots in Twitter," *Soc. Netw. Anal. Min.*, vol. 11, no. 1, p. 84, Sep. 2021, doi: 10.1007/s13278-021-00786-4.

[44] E. Alothali, K. Hayawi, and H. Alashwal, "Characteristics of Similar-Context Trending Hashtags in Twitter: A Case Study," in *Web Services – ICWS 2020*, Cham, 2020, pp. 150–163. doi: 10.1007/978-3-030-59618-7_10.